\documentclass[reprint,groupedaddress,aps,pra,groupedaddress,amsmath,amssymb,showpacs]{revtex4-1}
\usepackage{booktabs,ulem}
\usepackage[colorlinks,linkcolor=blue,anchorcolor=blue,citecolor=blue,
bookmarksnumbered]{hyperref}
\usepackage{graphicx,amsmath,amssymb,epsfig,color,epstopdf,float}
\usepackage{dcolumn,comment}
\usepackage{bm}
\def\be{\begin{equation}}
\def\ee{\end{equation}}
\def\bea{\begin{eqnarray}}
\def\eea{\end{eqnarray}}
\def\bes{\begin{subequations}}
\def\ees{\end{subequations}}
\def\a{\alpha}
\def\b{\beta}

\allowdisplaybreaks
\begin{document}
\title{Quantum Squeezing of Slow-Light Solitons}
\author{Jinzhong Zhu$^1$, Qi Zhang$^1$, and Guoxiang Huang$^{1,2,3}$}
\affiliation{$^1$State Key Laboratory of Precision Spectroscopy,East China Normal University, Shanghai 200062, China\\
             $^2$NYU-ECNU Joint Institute of Physics,  New York University Shanghai, Shanghai 200062, China\\
             $^3$Collaborative Innovation Center of Extreme Optics, Shanxi University, Taiyuan 030006, China
             }

\begin{abstract}
We investigate the quantum squeezing of slow-light solitons generated in a $\Lambda$-shaped three-level atomic system working under condition of electromagnetically induced transparency (EIT). Starting from the Heisenberg-Langevin and Maxwell equations governing the quantum dynamics of atoms and probe laser field, we derive a quantum nonlinear Schr\"{o}dinger equation controlling the evolution of the probe-field envelope. By using a direct perturbation approach to diagonalize the effective Hamiltonian (where the atomic variables have been eliminated), we carry out a detailed calculation on the quantum fluctuations of a slow-light soliton, expanded as a superposition of the complete and orthonormalized set of eigenfunctions obtained by solving the Bogoliubov–de Gennes (BdG) equations describing the quantum fluctuations. We show that due to the giant Kerr nonlinearity contributed from the EIT effect, significant quantum squeezing of the slow-light soliton can be realized within a short propagation distance. The results reported here are helpful for understanding the quantum property of slow-light solitons and for realizing light squeezing via EIT in cold atomic gases experimentally.

\end{abstract}


\maketitle

\section{Introduction}

Solitons, fascinating nonlinear wavepackets localized in space and time, can form in nonlinear media and propagate over long distances without distortion~\cite{Ablowitz2003}. The underlying physical mechanics for the formation and stability of solitons in conservative nonlinear systems is the exact balance between nonlinearity and dispersion (and/or diffraction).
Among various solitons studied so far, optical solitons have attracted much attention and have been investigated extensively because of their important applications in optical information processing and transmission~~\cite{Kivshar2003}.

Up to now, most of the methods for generating optical solitons are based on the excitations in far-off-resonance media (such as optical fibers). Since such media have generally very small Kerr nonlinearities, to excite the solitons intense electromagnetic radiations are needed. As a consequence, the optical solitons generated in this way travel with speeds closed to $c$ (the light speed in free space) and the media with large propagation lengths
are required~\cite{Haus1996,Kivshar2003,Agrawal2019}, which is not desirable for the optical information processing with devices of small sizes.

In recent years, many efforts have been paid to the research of electromagnetically induced transparency (EIT), a typical quantum interference effect occurring in resonant atomic gases, by which the light absorption due to the resonance between optical fields and atoms can be largely suppressed and giant Kerr nonlinearities can be acquired simultaneously~\cite{Fleischhauer2005,Khurgin2009}. Based on these important properties, it has been shown that ultraslow optical solitons (i.e. their propagating velocity are much less than $c$; see Sec.~\ref{section3c} below) ~\cite{Deng2004PRL,Deng2004OL,Huang2005PRE,Hang2006,Michinel2006,Qi2011,
Khadka2014,Facao2015,Shou2019} and their storage and retrieval~\cite{Chen2014,Bai2019} can also be realized. However, up to now all the  studies on slow-light solitons in EIT-based media have been limited
in semicalssical regime, i.e. the solitons obtained are described by classical optical fields. Comparing with the case in optical fibers~\cite{Haus1996,Kivshar2003,Agrawal2019} where various quantum effects of optical solitons were intensively studied both theoretically and experimentally in past decades~\cite{DrummondPRL1987,DrummondJOSAB1987,Potasek1988,HausPRA19891,
HausPRA19892,
HausJOSAB1990,YLai1990,Rosenbluh1991,YLai1993,Drummond1993,YLai1993,Lai1993IEEE,
YLai1995,Duan1995,YaoPRA2,Hagelstein1996,Haus1997c,Haus1998,
Haus2000,Matsko2000,Drummond2001,Fiorentino2002,Kozlov2002,Kozlov2003,
RKLee2004,Rand2005,Tsang2006,Tran2011,Honarasa2011,Drummond2014,Andersen2016,
Hosaka2016},
the quantum effect of slow-light solitons has never been considered up to now. Because slow-light solitons are produced at very low light levels (with power around ten microwatts; see Sec.~\ref{section3c}), their quantum effects (especially quantum squeezing) should be more significant than that of conventional solitons (e.g. solitons in optical fibers). Thus it is necessary to consider quantum effects of the slow-light solitons and reveal their specific quantum characters.

In this work, we present a quantum theory of slow-light solitons, which are generated in a resonant $\Lambda$-shaped three-level atomic gas working under condition of EIT. Based on the Heisenberg-Langevin and Maxwell (HLM) equations governing the quantum dynamics of the atomic gas and a probe laser field, we derive a quantum nonlinear Schr\"{o}dinger (QNLS) equation that controls the time evolution of the probe-field envelope. We construct an effective Hamiltonian and hence the effective field theory in which only the degrees of the optical field are involved.

By using a direct perturbation approach to diagonalize the effective Hamiltonian, we carry out detailed calculations on the quantum fluctuations of a slow-light soliton. Different from the approaches on quantum solitons in optical fibers~\cite{DrummondPRL1987,DrummondJOSAB1987,Potasek1988,HausPRA19891,HausPRA19892,
HausJOSAB1990,YLai1990,Rosenbluh1991,YLai1993,Drummond1993,YLai1993,Lai1993IEEE,
YLai1995,Duan1995,YaoPRA2,Hagelstein1996,Haus1997c,Haus1998,Haus2000,Matsko2000,Drummond2001,Fiorentino2002,Kozlov2002,Kozlov2003,
RKLee2004,Rand2005,Tsang2006,Tran2011,Honarasa2011,Drummond2014,Andersen2016,Hosaka2016}, here we expand the  quantum fluctuations of the soliton as a linear superposition of the complete and orthonormalized eigenfunction set, which are obtained by solving the BdG equations describing the quantum fluctuations. We show that, due to the giant Kerr nonlinearity contributed from the EIT effect in the atomic gas, comparing with optical-fiber solitons a significant quadrature squeezing of the slow-light soliton can be realized within a short propagation distance both classically (i.e. squeezed soliton width) and quantum mechanically (i.e. squeezed quantum fluctuations); additionally, in company with the slow-light soliton squeezing, an atomic spin squeezing can also be realized in the system. The research results reported here are useful for understanding the quantum property of slow-light solitons, for developing the quantum theory of nonlinear optics~\cite{Drummond2014}, and for possible applications in optical quantum information processing and precision measurements.

The remainder of the article is arranged as follows. In Sec.~\ref{Sec2}, a description of the model under study is presented.  In Sec.~\ref{Sec3}, the QNLS equation and the effective Hamilton describing the dynamics of the probe field envelope are derived; the slow-light solitons under coherent-state approximation
and their classical squeezing of temporal width are discussed. In Sec.~\ref{Sec4}, a direct perturbation approach is applied to solve the QNLS equation and the quantum squeezing of slow-light solitons is studied in detail. Lastly, Sec.~\ref{Sec5} summarizes the main results obtained in this work.

\section{Model}\label{Sec2}

\begin{figure}
\includegraphics[width=1\columnwidth]{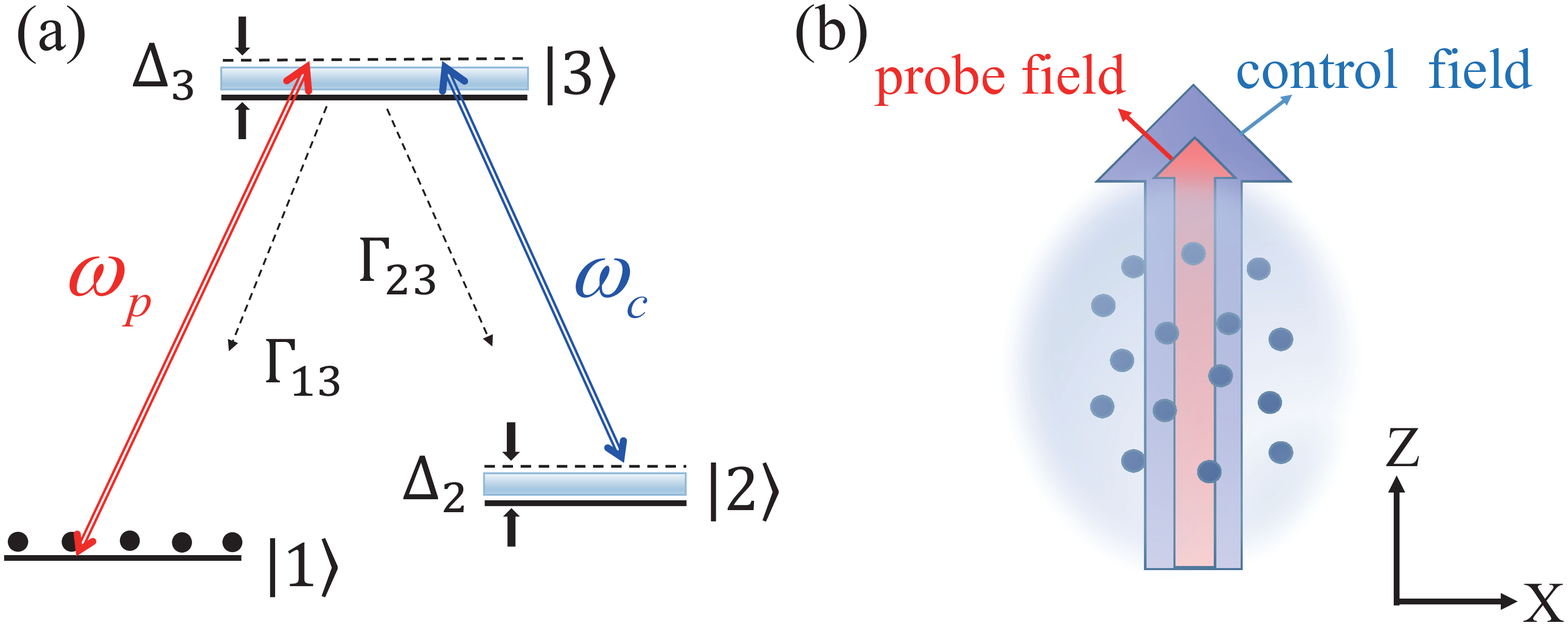}
\caption{(a)~Energy-level diagram and excitation scheme of the EIT-based $\Lambda$-type atomic gas. The pulsed probe field of center angular frequency $\omega_p$ couples the transition $|1\rangle\leftrightarrow|3\rangle$; the continuous-wave control field of center angular frequency $\omega_c$ couples the transition $|2\rangle\leftrightarrow|3\rangle$.
$\Delta_{2}$ ($\Delta_{3}$) is two-photon (one-photon) detuning; $\Gamma_{\alpha 3}$ is the decay rate of the spontaneous emission from the state $|3\rangle$ to $|\alpha\rangle$ ($\alpha=1,2$). Solid black dots mean that the atoms are initially prepared at the ground state $|1\rangle$.
(b)~Possible experimental geometry. To reduce Doppler effect, both probe and control fields are assumed to propagate along $z$ direction. }
\label{Fig1}
\end{figure}

We start to consider a cigar-shaped cold atomic gas with a $\Lambda$-shaped three-level configuration, composed of two nearly degenerate ground states $|1\rangle$ and $|2\rangle$ and an excited state $|3\rangle$, as schematically shown in Fig.~\ref{Fig1}.
A pulsed probe laser field (with center angular frequency $\omega_p$ and wavenumber $k_p$) couples the transition $|1\rangle\leftrightarrow|3\rangle$; a continuous-wave control laser field (with center angular frequency $\omega_c$ and central wavenumber $k_c$) couples  the transition $|2\rangle\leftrightarrow|3\rangle$.
$\Delta_{2}=\omega_{p}-\omega_{c}-(\omega_{2}-\omega_{1})$ and $\Delta_{3}=\omega_{p}-(\omega_{3}-\omega_{1})$ are respectively two- and one-photon detunings, with $\hbar\omega_{\alpha}$ the eigenenergy of the state $|\alpha\rangle$; $\Gamma_{\alpha3}(\alpha=1,2)$ is the decay rate of the spontaneous emission from $|3\rangle$ to $|\alpha\rangle$.

In order to suppress Doppler effect, both probe and control fields are arranged to propagate along the same (i.e. $z$) direction. The cigar-shaped atomic gas can be 
taken to be filled in a waveguide. Thereby, a reduced (1+1)-dimensional  (i.e. time plus space in the $z$-axis) model is sufficient to describe the dynamics of the system, with the total electric field given by
\begin{subequations}\label{Efield}
\begin{align}
& {\hat{\bf E}}(z,t)={\hat{\bf E}}_{p}(z,t)+{\bf E}_{c}(z,t),\label{Efield1}\\
& {\hat{\bf E}}_{p}(z,t)={\bf e}_{p}{\cal E}_{p}\hat{E}_{p}(z,t)e^{i(k_{p}z-\omega_{p})t}+{\rm h.c.}, \label{Efield2}\\
& {\bf E}_{c}(z,t)={\bf e}_{c}{\cal E}_{c}(z,t)e^{i(k_{c}z-\omega_{c})t}+{\rm c.c.}. \label{Efield3}
\end{align}
\end{subequations}
Here h.c. (c.c.) denotes Hermitian (complex) conjugate;
${\hat{\bf E}}_{p}$ is the quantized probe field, with ${\bf e}_{p}$ the  unit polarization vector, ${\cal E}_{p}\equiv\sqrt{\hbar\omega_{p}/(2\varepsilon_{0}V)}$ the field amplitude of single-photon, and $V$ the quantization volume; ${\bf E}_{c}$ is the control field with unit polarization vector ${\bf e}_{c}$ and field amplitude ${\cal E}_{c}$, which is assumed to be strong enough and thus can be taken as a classical and undepleted one. The annihilation operator of probe photons, $\hat{E}_{p}(z,t)$, is a slowly-varying function of $z$ and $t$ and  obeys the equal-time commutation relation $[\hat{E}_{p}(z,t),\hat{E}_{p}^{\dag}(z',t)]=L\delta(z'-z)$, where $L$ is the quantization system length along the $z$-axis.

Under electric-dipole, rotating-wave, and paraxial approximations, the Hamiltonian of the system reads
\begin{align}\label{Hamiltonian}
{\hat H}&=\int dz\bigg[-\frac{\hbar c}{L}\hat{E}_{p}^{\dag}\left(i\frac{\partial}{\partial z}\right)\hat{E}_{p} -\frac{\hbar N}{L}\left(\sum_{\alpha=2,3}\Delta_{\alpha}\hat{S}_{\alpha\alpha}\right.
\notag \\
&\hspace{4mm}\left.+g_{p}\hat{S}_{31}^\dag\hat{E}_{p}+\Omega_c\hat{S}_{32}^\dag+{\rm h.c.}\right)\bigg].
\end{align}
Here $N$ is the total atomic number of the system;
\be \label{Sab}
\hat{S}_{\alpha\beta}(z,t)=
\hat{\sigma}_{\beta\alpha}\,e^{i[(k_\beta-k_\alpha)z-(\omega_\beta-\omega_\alpha
+\Delta_\beta-\Delta_\alpha)t]}
\ee
are atomic transition operators from the states $|\alpha\rangle$ to $|\beta\rangle$ $(\alpha,\beta = 1,2,3)$, obeying the equal-time commutation relation $\left[\hat{S}_{\alpha\beta}(z,t),\hat{S}_{\mu\nu}(z',t) \right]=(L/N)\delta(z-z')[\delta_{\alpha\nu}\hat{S}_{\mu \beta}(z,t)-\delta_{\mu \beta}\hat{S}_{\alpha\nu}(z,t)]$; here $k_1=0$, $k_2=k_p-k_c$, and $k_3=k_p$; $\Omega_c=(\mathbf{e}_c\cdot \mathbf{p}_{32})\mathcal{E}_c/\hbar$ is the half Rabi frequency of the control field; $g_{p}=(\mathbf{e}_p\cdot\mathbf{p}_{31})\mathcal{E}_p/\hbar$ is the coefficient denoting the strength of the coupling between the probe photon and the transition $|1\rangle\leftrightarrow|3\rangle$ (i.e. single-photon half Rabi frequency);  $\mathbf{p}_{\alpha\beta}$ is the electric dipole matrix element associated with the transition from $|\beta\rangle$ to $|\alpha\rangle$.

The dynamics of the system is governed by the Heisenberg-Langevin and the Maxwell (HLM) equations, given by
\begin{subequations}\label{HLM}
\begin{align}
& \frac{\partial}{\partial t}{\hat S}_{\alpha\beta}=-i\left[{\hat S}_{\alpha\beta},\frac{{\hat H}}{\hbar}\right]-{\hat{\cal L}}({\hat S}_{\alpha\beta})+{\hat F}_{\alpha\beta}, \label{HLM1}\\
& i\left(\frac{\partial}{\partial z}+\frac{1}{c}\frac{\partial}{\partial t}\right)\hat{E}_{p}+\frac{g_{p}^\ast N}{c}{\hat S}_{31}=0, \label{HLM2}
\end{align}
\end{subequations}
where ${\hat{\cal L}}({\hat{S}_{\alpha\beta}})$ is the $3\times3$ relaxation matrix including the atomic decay rates of the spontaneous emission and dephasing, ${\hat F}_{\alpha\beta}$ are $\delta$-correlated Langevin noise operators introduced to preserve the Heisenberg commutation relations for the operators of the atoms and the probe field. Explicit expressions of Eq.~(\ref{HLM}a) are presented in Appendix~\ref{app1}.

The model described above can be realized by many atomic systems. One of the candidates is the laser-cooled alkali $^{87}$Rb gas, with the levels chosen to be $|1\rangle=|5^2S_{1/2},F=1,m_{F}=1\rangle$, $|2\rangle=|5^2S_{1/2},F=2,m_{F}=1\rangle$ and $|3\rangle=|5^2P_{3/2},F=2,m_{F}=1\rangle$. System parameters (used in the following calculations) are given by $\Gamma_{13}=\Gamma_{23}=2\pi\times3\,{\rm MHz}$~\cite{Steck}.
%

\section{Quantum nonlinear Schr\"{o}dinger equation and effective Hamiltonian}\label{Sec3}

\subsection{Quantum NLS equation}\label{section3a}

We are interested in the nonlinear quantum dynamics of the system, which needs to solve the quantum field equations (\ref{HLM1}) and (\ref{HLM2}) that are quantized and nonlinearly coupled each other. It is desirable, under some approximations, to reduce such equations to an effective one by which the atomic degrees of freedom are eliminated. Such effective field theory approach~\cite{Larre2015,Larre2016} has been recently used to the study of Rydberg polaritons in atomic gases~\cite{Gullans2016}. Based on such a spirit, here we give a (heuristic) derivation on the QNLS equation describing the nonlinear evolution of the probe-field envelope $\hat{E}_{p}$.

The derivation is divided into two steps. The first step is to neglect the nonlinearity in the HLM equations  (\ref{HLM1}) and (\ref{HLM2}) and consider only the linear propagation of the quantized probe field. Due to the coupling with the atoms, the probe field displays dispersion during propagation. We assume that the bandwidth of the probe pulse is not too narrow (e.g. the order of the 10\,MHz considered here), thus it is enough to include the dispersion to the second order (i.e. group-velocity dispersion). Then from linearized Eqs.~(\ref{HLM1}) and (\ref{HLM2}) we can obtain a linear Schr\"odinger equation after eliminating the atomic variables, i.e. Eq.~(\ref{Linear Eq}) given in Appendix~\ref{app0}.

The second step is to neglect the dispersion in the HLM equations  (\ref{HLM1}) and (\ref{HLM2}) and consider only the photon-photon interaction in the probe field intermediated by the atoms.  We assume that the light intensity of the probe pulse is weak but the photon-photon interaction cannot be neglected, thus it is necessary to include the lowest-order Kerr effect. Then by neglecting the dispersion in Eqs.~(\ref{HLM1}) and (\ref{HLM2}) we can obtain a nonlinear equation on $\hat{E}_{p}$ after eliminating the atomic variables, i.e. Eq.~(\ref{Nonlinear Eq}) given in the Appendix~\ref{app0}.

To obtain the envelope equation of the probe field valid to the approximations of the second-order dispersion and cubic nonlinearity, we combine Eq.~(\ref{Linear Eq}) and Eq.~(\ref{Nonlinear Eq}), which gives the following QNLS equation
\begin{align}\label{QNLS0}
& i\left[\left(\frac{\partial}{\partial z}+\frac{1}{V_{g}}\frac{\partial}{\partial t}\right)+{\rm Im}(K_{0})\right]\hat{E}_{p}-\frac{K_{2}}{2}\frac{\partial^2}{\partial t^2}\hat{E}_{p} \notag\\
& +W|g_p|^{2}\hat{E}_{p}^{\dag}\hat{E}_{p}\hat{E}_{p}-i{\hat{\cal F}}_{p}e^{-i{\rm Re}(K_{0})z}=0.
\end{align}
 Here $K_{0}\equiv K|_{\omega=0}$, $V_{g}^{-1}\equiv K_{1}\equiv(\partial K/\partial\omega)|_{\omega=0}$ is the group-velocity of the  probe field, $K_{2}\equiv(\partial^2K/\partial\omega^2)|_{\omega=0}$ is the coefficient of the group-velocity dispersion, with $K(\omega)$ the linear dispersion relation of the system.  $W=\hbar^2\omega_{p}\chi_{p}^{(3)}/(2 c|\mathbf{e}_{p}\cdot\mathbf{p}_{31}|^{2})$ is the coefficient of third-order Kerr nonlinearity, which is proportional to the third-order nonlinear optical susceptibility $\chi_{p}^{(3)}$. ${\hat{\cal F}}_{p}(z,t)$ is the $\delta$-correlated induced Langevin noise operator and ``Re'' means taking real part. For the detailed derivations of the QNLS~(\ref{QNLS0}) and explicit expressions of $K(\omega)$,  $\chi_{p}^{(3)}$, and ${\hat{\cal F}}_{p}(z,t)$, are given in the Appendix~\ref{app0}.

Generally, the coefficients of QNLS equation~(\ref{QNLS0}) take complex values due to the near-resonant interaction character of system. However, under the condition of EIT, i.e. $|\Omega_c|^2\gg \gamma_{21}\gamma_{31}$, the imaginary part of these coefficients can be made much smaller than their corresponding real parts~\cite{Fleischhauer2005}. One can check this by considering a set of experimentally achievable parameters, given by
${\cal N}_a$ (atomic density)$=7.33\times10^{10}\,{\rm cm}^{-3}$, $|g_{p}|^2N/c=2\times10^{9}\,{\rm cm^{-1}s^{-1}}$, $\Omega_{c}=2\pi\times18\,{\rm MHz}$, $\Delta_{2}=2\pi\times1.2 \,{\rm MHz}$, $\Delta_{3}=2\pi\times60\,{\rm MHz}$, and $t_{0}\,\, {\rm (the\,\, pulse\,\, duration\,\, of\,\, the\,\, probe\,\, pulse)}=1.4\times10^{-7}\,{\rm s}$.
We obtain  $K_{1}=(2.59+0.074i)\times10^{-7}\,{\rm cm^{-1}s}$,
$K_{2}=(2.03+0.19i)\times10^{-14}\,{\rm cm^{-1}s^2}$, $W=-(4.48+0.086i)\times10^{-16}\,{\rm cm^{-1}s^2}$. Based on these results, we have
\begin{equation}\label{chi3}
\chi_p^{(3)}=-(6.57+0.13) \times 10^{-10}\,{\rm m}^2{\rm V}^{-2},
\end{equation}
and hence ${\rm sgn}[-{\rm Re}(W)/{\rm Re}(K_2)]=1$ (here sgn is sign function), which means that the system is self-focused, useful for generating bright solitons.
We see that the imaginary parts of these coefficients are indeed much smaller than their corresponding real parts. The physical reason for the small imaginary parts of the QNLS equation (\ref{QNLS0}) is due to the fact that the quantum destruction interference effect induced by the control field (i.e. EIT effect) brings a significant suppression of the spontaneous emission of the atoms, which results also in a giant Kerr nonlinearity in the system (with $|\chi_p^{(3)}|\approx  10^{-10}\,{\rm m}^2{\rm V}^{-2}$)
that is several orders larger than that of optical fibers (with $|\chi_p^{(3)}|\approx 10^{-19}\,{\rm m^2/V^2}$~\cite{Agrawal2019}).

The optical depth is a useful parameter characterizing the coupling between the light field and the atoms, defined by ${\rm OD}\equiv\omega_{p}|{\bf p}_{31}|^2{\cal N}_{a}L/(2\varepsilon_{0}c\hbar\gamma_{31})$, which contains the product of ${\cal N}_{a}$ and $L$ (the length of the atomic cell).
The OD in cold atmic gases up to 300 was achieved by Vernaz-Gris {\it et al.} (where $L=2.5$ cm) and by Wang {\it et al.} (where $L=3$ cm)\cite{Vernaz2018,Wang2019}. Here we take $L=2~{\rm cm}$, then the optical depth in our system is OD$\approx 212$,  which is less than 300 and hence can be realized in current-day experiments.

After neglecting the imaginary parts of $K_1$, $K_2$, and $W$,
Eq.~(\ref{QNLS0}) can be written as the dimensionless form
\begin{equation}\label{QNLSE1}
i\frac{\partial}{\partial s}\hat{U}+\frac{\partial^2}{\partial\tau^2}\hat{U}+2g\hat{U}^{\dag}\hat{U}\hat{U}
=-2i\nu\hat{U}+i\hat{f}_{p},
\end{equation}
with $\hat{U}=\hat{E}_{p}/\sqrt{n_{0}}$ ($n_{0}\gg 1$ is typical mean photon number in the probe field), $s=z/(2L_{\rm disp})$, $\tau=(t-z/V_{g})/t_{0}$, ${\hat f}_{p}=2L_{\rm disp}{\hat{\cal F}}_{p}e^{-i{\rm Re}(K_{0})z}$, $\nu=L_{\rm disp}/L_{\rm abs}$, $g= L_{\rm disp}/L_{\rm nln}$. Here  $L_{\rm disp}\equiv t_{0}^2/|{\rm Re}(K_{2})|$, $L_{\rm nln}\equiv [n_{0} |g_{p}|^2|{\rm Re}(W)|]^{-1}$, and $L_{\rm abs}\equiv 1/{\rm Im}(K_{0})$ are typical dispersion length, nonlinearity length, and absorption length of the probe field, respectively~\cite{note}. The commutation relation for ${\hat f}_{p}$ has the form~\cite{YLai1995}
\be
[{\hat f}_{p}(s,\tau),{\hat f}_{p}^\dag(s',\tau')]=4\nu\delta(s-s')\delta(\tau-\tau').
\ee

Making the transformation $\hat{U}=\hat{{\bar U}}e^{-i\mu s}$,
Eq.~(\ref{QNLSE1}) is converted into the form
\begin{eqnarray}\label{QNLSE2}
i\frac{\partial}{\partial s}\hat{\bar{U}}=
 -\frac{\partial^2}{\partial\tau^2}\hat{\bar{U}}-2g\hat{\bar U}^{\dag}\hat{\bar{U}}\hat{\bar{U}}-\mu\hat{\bar{U}}
 -2i\nu\hat{\bar U}+i\hat{\bar f}_{p},
\end{eqnarray}
where $\hat{\bar f}_{p}=\hat{f}_{p}e^{i \mu s}=2L_{\rm disp}\hat{{\cal F}}_p(s,\tau)\exp \{i[\mu-2{\rm Re}(K_0)L_{\rm disp}]s\}$, with the  parameter $\mu$ the ``chemical potential'' to be specified lately. In Eq.~(\ref{QNLSE1}) and Eq.~(\ref{QNLSE2}), the non-dimensional parameter $g$ characterizes the magnitude of the Kerr nonlinearity in the system.

\subsection{Effective Hamiltonian and equal-space commutation relation}

The QNLS equation (\ref{QNLSE2}) governs the quantum evolution of the probe-field envelope under the condition of EIT; it is obtained by the elimination of the atomic variables. For convenience for the study of the photon dynamics in the system, one can adopt another viewpoint based on effective field theory mentioned above. The QNLS equation~(\ref{QNLSE2}), after neglecting the damping and noise terms (see the reason given below), can be taken to describe a reductive physical system, controlled  by the effective Hamiltonian
\begin{equation}\label{Heff0}
\hat{H}_{\rm eff}=\int_{-\infty}^{+\infty}d\tau{\hat{\bar U}}^{\dagger}
  \left(-\frac{\partial^2}{\partial\tau^2}-\mu-g{\hat{\bar U}}^{\dag}{\hat{\bar U}}\right){\hat{\bar U}}.
\end{equation}
If the following equal-$s$ commutation relation
\begin{equation}\label{CR}
[\hat{\bar U}(s,\tau),\hat{\bar U}^\dag(s,\tau')]=\delta(\tau-\tau'),
\end{equation}
is assigned, the Heisenberg equation of motion
\begin{equation}\label{HeisEqu}
i\frac{\partial}{\partial s}{\hat{\bar U}}=\left[{\hat{\bar U}},{\hat H}_{\rm eff}\right],
\end{equation}
yields the QNLS equation (\ref{QNLSE2}). In this way, the problem of photon dynamics can be studied based on Eqs.~(\ref{Heff0})-(\ref{HeisEqu}), in which the role of time and distance are exchanged in comparison with conventional quantum approaches. Such effective theory has been widely adopted in recent years for the study on photon propagation in optical fibers, bulk Kerr medium, Rydberg atoms, and other physical systems~\cite{HausPRA19891,HausPRA19892,Matsko2000,Larre2015,Gullans2016,Larre2016}.

When writing the effective Hamiltonian, the last two terms on the right-hand side of Eq.~(\ref{QNLSE2}) have been neglected, which is approximately valid based on the following reasons. (i)~Based on the physical parameters given in Sec~\ref{section3a}, we obtain~$L_{\rm disp}=0.97~{\rm cm}$, $L_{\rm abs}=46.19~{\rm cm}$. Thus the dimensionless absorption coefficient $\nu=L_{\rm disp}/L_{\rm abs}=2.09\times10^{-2}\ll1$, which means that the damping term $-2i\nu\hat{\bar U}$ is very small and can be disregarded. (ii)~When considering the thermal reservoir coupled with the atomic gas, the two-time correlation functions for the induced Langevin noise operator ${\hat {\bar f}}_{p}$ are given by $\langle{\hat {\bar f}}_{p}(s,\tau){\hat {\bar f}}_{p}^\dag(s',\tau')\rangle=4\nu({\bar n}_{\rm th}+1)\delta(s-s')\delta(\tau-\tau')$ and $\langle{\hat {\bar f}}_{p}^\dag(s,\tau){\hat {\bar f}}_{p}(s',\tau')\rangle=4\nu {\bar n}_{\rm th}\delta(s-s')\delta(\tau-\tau')$. Here $\langle\cdots\rangle$ is the expectation value by taking the trace over the reservoir variables; ${\bar n}_{\rm th}=\{\exp[\hbar\omega_{p}/(k_{\rm B}T)]-1\}^{-1}$ is the mean photon number in the thermal reservoir, with $k_{\rm B}$ the Boltzmann constant and $T$ the temperature of the reservoir. Since at optical frequencies and in the ultracold environment, one has $\hbar\omega_{p}\gg k_{\rm B}T$ which yields ${\bar n}_{\rm th}\approx 0$. 
Indeed, due to the EIT effect one has $\langle\hat{S}_{33}\rangle\approx0$, which means Langevin noise operators make no contribution to the normally-ordered correlation functions of system operators~\cite{Gorshkov20071,Gorshkov20072}.

\subsection{Slow-light solitons}\label{section3c}

We now investigate the quantum nonlinear dynamics based on the description given by Eqs.~(\ref{Heff0})-(\ref{HeisEqu}). We first consider the classical limit of such a reduced system, which is valid when the probe field is in a coherent state, i.e. the operator $\hat{\bar U}$ can be replaced by a c-number function $U_0$.  Then the Heisenberg equation (\ref{HeisEqu}) becomes a classical NLS equation of the form $i\partial  U_0/\partial s+\partial^2 U_0/\partial\tau^2+2g|U_0|^2 U_0+\mu U_0=0$, which admits the fundamental bright-soliton solution
\begin{align}\label{CSoliton0}
U_0(s,\tau)&=\frac{A_{0}\sqrt{g}}{2}{\rm sech}\left[\frac{A_{0}g}{2}(\tau-\tau_{0}-2p_{0}s)\right] \notag \\
& \hspace{3mm}\times \exp\left[ip_{0} (\tau-\tau_{0})-ip_{0}^2 s+i\theta_{0}\right],
\end{align}
with $\mu=-A_{0}^{2}g^{2}/4$.
Here $A_{0}$, $\theta_{0}$, $p_{0}$, and $\tau_{0}$ are real parameters, related to the soliton amplitude (defined by $A_{0}\equiv\int_{-\infty}^{+\infty}  d\tau |U_0|^2= n_{0}^{-1}\int_{-\infty}^{+\infty}d\tau |E_p|^2$), initial phase, ``momentum'', and initial ``position'', respectively~\cite{HausJOSAB1990}.

By using the physical parameters given in Sec.~\ref{section3a}, we obtain $V_{g}\approx 1.28 \times 10^{-4} c$. The propagating velocity $V_s$ of the soliton has a small modification from $V_{g}$. For example, for $p_0=1$
\begin{align}\label{Vs}
V_s=\frac{V_{g}}{1+p_{0}\frac{V_{g}t_{0}}{L_{\rm disp}}}\approx8.23\times 10^{-5}c,
\end{align}
i.e. $V_s\approx V_{g}$ and both of them are much smaller than $c$. The significant slowdown of the optical pulses in the system is contributed by the EIT effect induced by the control field.

The threshold power of the probe field can be estimated by calculating the Poynting’s vector, given by ${P}_{\max}=2 \varepsilon_{0}c n_{p}S\hbar^2|{\rm Re}(K_2)|/(|{\bf e}_{p}\cdot{\bf p}_{31}|^2 t_0^2|{\rm Re}(W)|)$ \cite{Newell1992,Bai2019}. Here $n_p=1+c |\rm Re(K_0)|/\omega_p$  is the refractive index, $S=\pi R_0^{2}$ is the cross-section area ($R_0$ is the transverse radius) of the probe field. Using the above parameters and taking $R_0=200 ~\mu\mathrm{m}$, we obtain
\begin{align}\label{power}
{P}_{\max }\approx6.66~\mu\mathrm{W}.
\end{align}
Thus to generate the slow-light soliton a very low input power is needed, which is also contributed by the EIT effect that brings a giant enhancement of the Kerr nonlinearity in the system~\cite{Deng2004PRL,Huang2005PRE}. This is very different from the solitons generated in optical fibers where, due to small Kerr nonlinearity, much larger light power and hence long propagation distance are required~\cite{Haus1996,Kivshar2003,Agrawal2019}.

\begin{figure}
\centering
\includegraphics[width=1\columnwidth]{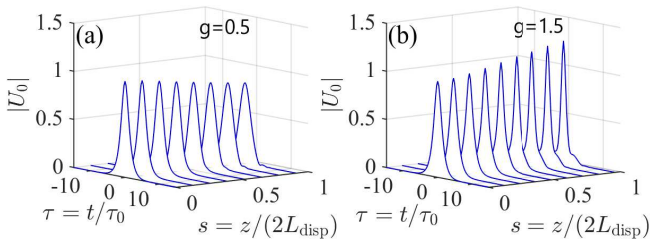}
\caption{Classical squeezing (compression) of the width of slow-light soliton by different Kerr nonlinearities (characterized by the parameter g).
(a)~$|U_0|$ as a function of $\tau=t/\tau_0$ and $s=z/(2L_{\rm disp})$, with $g=0.5$ and the input pulse $U_0(0,\tau)={\rm sech}(\tau)$.
(b)~The same as (a) but with $g=1.5$.
}
\label{Fig2}
\end{figure}

For completeness, Fig.~\ref{Fig2} shows
the waveshape of the slow-light soliton  during propagation, obtained by taking $|U_0|$ as a function of $\tau=t/\tau_0$ and $s=z/(2L_{\rm disp})$, with the input pulse given by $U_0(0,\tau)={\rm sech}(\tau)$; panel (a) [panel (b)\,] is for the case of $g=0.5$  ($g=1.5$). We see that the soliton width in (b) (in which Kerr nonlinearity is stronger) is narrower than in (a) (in which Kerr nonlinearity is weaker). Hence in the classical description the Kerr nonlinearity of the system can be employed to make the soliton width squeezed (compressed).

\section{Diagonalization of the effective Hamiltonian and quantum squeezing of slow-light solitons}\label{Sec4}

\subsection{Diagonalization of the effective Hamiltonian}

Now we turn to investigate the quantum correction of the slow-light soliton $U_{0}$ in the system. We assume that the mean photon number $n_0$ in the probe field is large,
the quantum fluctuations of the soliton are weaker, so that the dimensionless probe field can take the form (Bogoliubov decomposition)~\cite{Bogolubov}
\begin{equation}\label{BD}
\hat{{\bar U}}(s,\tau)=U_{0}(\tau)+\hat{U}_{1}(s,\tau).
\end{equation}
Here $U_{0}(\tau)=U_{0}(s=0,\tau)$; $\hat{U}_{1}$ is the annihilation operator of photons in the probe field characterizing the quantum fluctuations on the soliton background $U_{0}(\tau)$. From (\ref{CR}) one can obtain the commutation relation
\be
[\hat{U}_{1}(s,\tau),\hat{U}_{1}^{\dag}(s,\tau')]
=\delta(\tau-\tau').
\ee

By using the ansatz (\ref{BD}) and neglecting the high-order terms of $\hat{U}_{1}$, the effective Hamiltonian~(\ref{Heff0}) is converted into a quadratic form of $\hat{U}_{1}$:
\begin{subequations}\label{Heff2}
\begin{align}
{\hat H}_{\rm eff}& =H_{0}+{\hat H}_{2}, \label{Heff21}\\
H_{0}& =\int_{-\infty}^{\infty} d\tau U_{0}(-\partial^2/\partial\tau^2-\mu-gU_{0}^2)U_{0},\label{Heff22}\\
{\hat H}_{2}& =\int_{-\infty}^{\infty} d\tau[\hat{U}_{1}^\dag(-\partial^2/\partial\tau^2-\mu\notag\\
&\hspace{0.4cm}-4gU_{0}^2)\,\hat{U}_{1}-gU_{0}^2(\hat{U}_{1}\hat{U}_{1}
+\hat{U}_{1}^\dag\hat{U}_{1}^\dag)],\label{Heff23}
\end{align}
\end{subequations}
Here $U_{0}$ obeys the equation $(-\partial^2/\partial\tau^2-\mu-2gU_{0}^2)U_{0}=0$. It gives the single soliton solution of  $U_{0}(\tau)=(A_{0}\sqrt{g}/2){\rm sech}[A_{0}g(\tau-\tau_{0})/2]$, which can be obtained from (\ref{CSoliton0}) by setting $s=0$ and $\mu=-A_{0}^{2}g^{2}/4$. Hence, by carrying out the integration in (\ref{Heff22}), we obtain $H_{0}=A_{0}^3g^2/6$.

Different from previous studies on quantum solitons in optical fibers~\cite{DrummondPRL1987,DrummondJOSAB1987,Potasek1988,HausPRA19891,
HausPRA19892,HausJOSAB1990,YLai1990,Rosenbluh1991,YLai1993,Drummond1993,YLai1993,
Lai1993IEEE,YLai1995,Duan1995,YaoPRA2,Hagelstein1996,Haus1997c,Haus1998,
Haus2000,Matsko2000,Drummond2001,Fiorentino2002,Kozlov2002,Kozlov2003,
RKLee2004,Rand2005,Tsang2006,Tran2011,Honarasa2011,Drummond2014,Andersen2016,
Hosaka2016}, our approach here is based on
searching a rigorous diagonalization of the effective Hamiltonian ${\hat H}_{\rm eff}$, which can provide a complete solution for all possible quantum fluctuations. To simplify our discussion, we use the new variables   $\sigma=A_{0}g(\tau-\tau_0)/2$, $\hat{U}_{1}=\sqrt{A_{0}g/2}{\hat w}$. Then the Hamiltonian for the quantum fluctuations, ${\hat H}_{2}$, can be written into the form
\begin{align}\label{H2}
{\hat H}_{2}=\frac{A_{0}^{2}g^{2}}{4}\int_{-\infty}^{+\infty}d\sigma\left[{\hat w}^\dag{\hat L}{\hat w}-{\rm sech}^2\,\sigma\,({\hat w}{\hat w}+{\hat w}^\dag{\hat w}^\dag)\right],
\end{align}
where ${\hat L}=-\partial^2/\partial\sigma^2-4{\rm sech}^2(\sigma)+1$, ${\hat w}$ satisfies the commutation relation $[{\hat w}(s,\sigma),{\hat w}^\dag(s,\sigma')]=\delta(\sigma-\sigma')$.

To obtain general solutions for the quantum fluctuation ${\hat w}$, a clear and standard way is to find a set of complete and orthogonal eigenfunctions that can be used to expand ${\hat w}$ as a linear superposition of these eigenfunctions. This is, however, equivalent to diagonalize the Hamiltonian (\ref{H2}) by using the following Bogoliubov canonical transformation
\begin{align}\label{w}
\nonumber {\hat w}(s,&\sigma)
=\sum_j\left[u_{j}(\sigma)\hat{a}_{j}(s)+v_{j}(\sigma)\hat{a}_{j}^\dag(s)\right] \\
&\hspace{2mm}+\int_{-\infty}^{+\infty}dk\left[u(\sigma,k)\hat{a}(s,k)+v(\sigma,k)
\hat{a}^\dag(s,k)\right].
\end{align}
Here the first term (summation) and the second term (integration) on the right-hand side are contributed respectively from the discrete and continuum spectra of the excitations created from the soliton background;
$u_j(\sigma)$, $v_j(\sigma)$, $u(\sigma,k)$, $v(\sigma,k)$ are the corresponding  eigen (or mode) functions; $\hat{a}_{j}(s)$,
$\hat{a}_{j}^{\dag}(s)$, $\hat{a}(s,k)$, and $\hat{a}^{\dag}(s,k)$ are the corresponding  annihilation operators of photons obeying the commutation relations $[\hat{a}_{j}(s),\hat{a}_{j'}^\dag(s)]=\delta_{jj'}$ and $[\hat{a}(s,k),\hat{a}^\dag(s,k')]=\delta(k-k')$  (with other commutators zero).

The key to diagonalize ${\hat H}_{2}$ is to find a complete set of the
eigenfunctions  $\{u_{j}(\sigma),u(\sigma,k),v_{j}(\sigma),v(\sigma,k)\}$, with which the explicit form of ${\hat w}$ can be determined. Following the
idea in Ref.~\cite{Huang2006}, we assume $u_{q}$ and $v_{q}$ satisfy
the BdG eigen equations~\cite{bdge}
\begin{subequations}\label{bdg}
\begin{align}
\hat{L}u_{q}(\sigma)-2{\rm sech}^{2}\sigma\nu_{q}(\sigma)&=-\lambda_{q}u_{q}(\sigma),\\
\hat{L}v_{q}(\sigma)-2{\rm sech}^{2}\sigma u_{q}(\sigma)&=\lambda_{q}v_{q}(\sigma),
\end{align}
\end{subequations}
where $q=j$ ($q=k$) is for the discrete (continuous) spectrum,
$u_k(\sigma)\equiv u(\sigma,k)$, $v_k(\sigma)\equiv v(\sigma,k)$. The above BdG equations have the following analytical solutions~\cite{Huang2006}
\begin{subequations}
\begin{align}
& u(\sigma,k)=-\frac{k^2+2ik\tanh(\sigma)-\tanh^2(\sigma)}{\sqrt{2\pi}(k^2+1)}e^{ik\sigma},\\
& v(\sigma,k)=-\frac{{\rm sech}^2(\sigma)}{\sqrt{2\pi}(k^2+1)}e^{ik\sigma},
\end{align}
\end{subequations}
for the continuum spectrum with eigen value $\lambda_{k}=-k^2-1$ ($-\infty<k<\infty$), and
\begin{subequations}
\begin{align}
    & u_{1}(\sigma)=\frac{2-\sigma\tanh(\sigma)}{2}{\rm sech}(\sigma),\\
    & u_{2}(\sigma)=\frac{{\tanh(\sigma)+\sigma}}{2}{\rm sech}(\sigma),\\
    & v_{1}(\sigma)=-\frac{\sigma\tanh(\sigma)}{2}{\rm sech}(\sigma), \\
    & v_{2}(\sigma)=\frac{{\tanh(\sigma)-\sigma}}{2}{\rm sech}(\sigma),
\end{align}
\end{subequations}
for the discrete spectrum with degenerate eigenvalue $\lambda_1=\lambda_2=0$.
It can be shown that all the eigenfunctions of both the continuous and the discrete spectra given above constitutes a complete set of eigenfunctions (i.e. Hilbert space), which means that they can be used to expand quantum fluctuations given by the form (\ref{w}). A detailed discussion on the above solutions is presented in Appendix~\ref{app3}.

Based on the above analytical results and substituting~(\ref{w}) into~(\ref{H2}), we can diagonalize $\hat{H}_2$ after carrying out the integrations over $\tau$ on the eigenfunctions. Then we obtain
\begin{align}\label{Heff_diag}
\nonumber {\hat H}_{\rm eff}&=\frac{A_{0}^3g^2}{6}+\frac{A_{0}^{2}g^{2}}{4}\left[{\hat P}_{2}^2-{\hat Q}_{1}^2\right.\\
&\hspace{0.5cm}\left.-\int_{-\infty}^{+\infty}dk\lambda_{k}\hat{a}^\dag(s,k)
\hat{a}(s,k)\right].
\end{align}
Here ${\hat Q}_{j}$ and  ${\hat P}_{j}$ $(j=1,2)$ are respectively
the coordinate operators and momentum operators related to the discrete-spectrum eigenfunctions, defined respectively by
\begin{subequations}\label{QP}
\begin{align}
& {\hat Q}_{j}=\frac{1}{\sqrt{2}}\left(\hat{a}_{j}+\hat{a}_{j}^\dag\right),\\
& {\hat P}_{j}=\frac{1}{\sqrt{2}i}\left(\hat{a}_{j}-\hat{a}_{j}^\dag\right),
\end{align}
\end{subequations}
which obey the commutation relation $[{\hat Q}_{j},{\hat P}_{j'}]=i\delta_{jj'}$.
In this way, the Hamiltonian of the system in the presence of the quantum fluctuations on the soliton background is diagonalized, which is useful for the study of quantum squeezing of the slow-light soliton.

\subsection{Quantum dynamics of slow-light solitons}

Based on the diagonalized effective Hamiltonian (\ref{Heff_diag}), we can easily examine the quantum dynamics of the slow-light soliton. The Heisenberg equation of motion for the operator $\hat{A}$ reads
$i\partial \hat{A}/\partial s =[\hat{A}, \hat{H}_{\rm eff}]$. Taking $\hat{A}$ to be $\hat{Q}_j(s)$, $\hat{P}_j(s)$, and $\hat{a}(s,k)$, we obtain the equations
\begin{subequations}\label{HE0}
\begin{align}
& \frac{\partial}{\partial s}\hat{Q}_{1}=0, \\
& \frac{\partial}{\partial s}\hat{P}_{1}-\frac{A_{0}^{2}g^{2}}{2}\hat{Q}_{1}=0, \label{xi} \\
&  \frac{\partial}{\partial s}\hat{P}_{2}=0,\\
& \frac{\partial}{\partial s}\hat{Q}_{2}-\frac{A_{0}^{2}g^{2}}{2}\hat{P}_{2}=0. \label{alpha},\\
& i\frac{\partial}{\partial s}\hat{a}(s,k)+\frac{A_{0}^{2}g^{2}}{4}\lambda_{k}\hat{a}(s,k)=0.
\end{align}
\end{subequations}
It is easy to get the exact solutions of these equations, given by
\bes \label{SolutionofHE}
\bea
&& \hat{Q}_{1}(s)=\hat{Q}_{1}(0), \\
&& \hat{P}_{1}(s)=(A_{0}^{2}g^{2}s/2)\hat{Q}_{1}(0)+\hat{P}_{1}(0),\\
&& \hat{P}_{2}(s)=\hat{P}_{2}(0),\\
&& \hat{Q}_{2}(s)=(A_{0}^{2}g^{2}s/2)\hat{P}_{2}(0)+\hat{Q}_{2}(0),\\
&& \hat{a}(s,k)=\hat{a}(0,k)\exp(iA_{0}^{2}g^{2}\lambda_{k}s/4).
\eea
\ees
where $\hat{Q}_{j}(0)$, $\hat{P}_{j}(0)$, $\hat{a}(0,k)$ are the values of $\hat{Q}_{j}(s)$, $\hat{P}_{j}(s)$, $\hat{a}(s,k)$ at $s=0$, respectively.

From Eqs.~(\ref{HE0}) and their solutions (\ref{SolutionofHE}), we have the following conclusions: (i)~The  quantum fluctuations contributed by the discrete spectrum display specific characters.
The position operator $\hat{Q}_{1}$ (momentum operator $\hat{P}_{2}$) remains unchanged, but it becomes correlated with the momentum operator ${\hat P}_{1}$ (position operator ${\hat Q}_{2}$) during propagation. Such correlations between $\hat{Q}_j$ and $\hat{P}_j$ ($j=1,2$) lead to phase diffusion and position spreading of the soliton, contributed by the Kerr nonlinearity (characterized by the nonlinear parameter $g$). (ii)~The quantum fluctuation for the continuum-mode $k$ has only a simple effect, i.e. a phase shift
to the same mode caused by the Kerr nonlinearity; no correlation between different modes occurs during propagation.

By using the results (\ref{CSoliton0}), (\ref{w}), and (\ref{QP}), we can obtain the approximate expression of the quantized probe field

\begin{align}\label{ueqs}
\hat{\bar{U}}\left(s,\tau\right)
&= \hat{A}(s,\tau) \operatorname{sech}\left[\frac{A_{0}g}{2}\tau-\frac{{\hat Q}_{2}(s)}{\sqrt{A_{0}}}\right]\notag\\
&\,\,\,\,\times\exp\left[i\frac{{\hat P}_{1}(s)}{\sqrt{A_{0}}}+i\frac{g\sqrt{A_{0}}}{2}{\hat P}_{2}(s)\tau\right],\\
\hat{A}(s,\tau)=& \frac{\sqrt{g}}{2}\left\{A_{0}+\frac{1}{\sqrt{A_0}}\left[1-\frac{ A_{0}g}{2}
\tau\right.\right.\nonumber\\
&\hspace{1cm}\times\left. \left.\tanh\left(\frac{A_{0}g}{2}\tau\right)\right]{\hat Q}_{1}(s)\right\}.
\end{align}
When obtaining this result, a renormalization technique~\cite{Nayeh} has been used, also the contribution of the quantum fluctuations from the continuous spectrum has been neglected because such fluctuations make much smaller contribution comparing those from the discrete spectrum and they spreads and depart rapidly during the propagation.

From the expression (\ref{ueqs}), one can see that the soliton displays various quantum fluctuations, which are contributed by the discrete spectrum and propagate together with the soliton. The quantum fluctuations can be divided into two categories. One is characterized by the conjugated operator pair $\hat{Q}_{1}$ and $\hat{P}_{1}$, describing respectively the amplitude
(photon number) and phase fluctuations; the other one is
characterized by the conjugated operator pair $\hat{Q}_{2}$ and $\hat{P}_{2}$, describing respectively the position and momentum fluctuations.

The above results can be used to obtain the numerical results of the quantum fluctuations of the soliton. Let $|\Psi\rangle =|n_{0},n_1,n_2,n_c\rangle$
denote the quantum state with $n_{0}$ photons in the soliton; $n_1$ and $n_2$
are photon numbers of the photons occupying the first and the second discrete modes, and $n_c$ is the number of photons in all continuous modes. We assume that at the entrance of the system ($s=0$) the quantum state of the probe field is prepared to be the ``vacuum'' state $|\Psi_0 \rangle=|n_{0},0,0,0\rangle$, i.e. the probe field has no quantum fluctuation. Based on the above analytical results, we obtain $\langle \hat{Q}_j(s)\rangle=\langle \hat{P}_j(s)\rangle=0$,
$\langle \hat{Q}_j^2(0)\rangle=\langle \hat{P}_j^2(0)\rangle=1/2$\, ($j=1,2$),
and the variances (mean-squared derivations) as functions of $s$ are given by
\begin{subequations}\label{unc}
\begin{align}
 &\langle{\hat Q}_{1}^2(s)\rangle=\langle{\hat P}_{2}^2(s)\rangle=\frac{1}{2},\\
 &\langle{\hat Q}_{2}^2(s)\rangle=\langle{\hat P}_{1}^2(s)\rangle=\frac{1}{2}\left(1+\frac{1}{4}g^4A_{0}^4s^2\right),
\end{align}
\end{subequations}
here $\langle\cdots\rangle \equiv\langle\Psi_0|\cdots|\Psi_0\rangle$. Note that, even there is no fluctuation before the soliton enters into the atomic medium, the quantum fluctuations will be generated in the probe field.  This is very different from the classical soliton system considered by Yan {\it et al.}~\cite{Yan1998PRE}.

\subsection{Quantum squeezing of slow-light solitons and its detection}\label{sec3D}

\subsubsection{Quantum squeezing of slow-light solitons}

The quantum squeezing of light have received a great deal of attention in recent years and found important applications in many fields, especially in quantum precision measurements (including the detection of weak forces such as gravitational waves)~\cite{Andersen2016,Schnabel2017}. Based on the results obtained above, we now explore the physical property of the quantum squeezing of the slow-light soliton in the present system.  Since the quantum fluctuations from continuous spectrum are much smaller compared to those from the discrete spectrum~\cite{HausJOSAB1990,YLai1993}, they will be disregarded in the following discussions.

For describing the quantum squeezing, we introduce the following quadrature operators related to $\hat{a}_{j}$ at the angle $\theta$~\cite{Mandel}
\bea \label{QuadOperators}
\hat{X}_{j,\theta}(s)
&& =\frac{1}{\sqrt{2}}\left[\hat{a}_j(s)\,e^{-i\theta}
+\hat{a}_j^{\dag}(s)\,e^{i\theta}\right]\label{QuadOperators1}\nonumber\\
&& =\hat{Q}_j(s)\cos\theta+\hat{P}_j (s)\sin\theta,\label{QuadOperators2}
\eea
which satisfies the commutation relation $[{\hat X}_{j,\theta},{\hat X}_{j',\theta}]=i\delta_{jj'}$ \,($j,j'=1,2$). When obtaining (\ref{QuadOperators2}) the definition (\ref{QP}) has been used.   $\hat{X}_{1,\theta}(s)$\,  [$\hat{X}_{2,\theta}(s)$] describes the quantum fluctuations of the amplitude and phase (position and momentum) of the soliton.

With this notations, the probe field can be expressed as a superposition of the soliton and the quantum fluctuations, i.e. $\hat{{\bf E}}_p={\bf E}_{\rm sol}+\hat{{\bf E}}_{\rm fluc}$. Using (\ref{Efield2}),
(\ref{w}), (\ref{QP}), (\ref{QuadOperators2}), and returning to the original variables, we obtain the expression of  soliton part
\bea
&& {\bf E}_{\rm sol}={\bf e}_p D_0 {\rm sech} \left[\frac{A_{0}g}{2t_{0}}\left(t-\frac{z}{V_{g}}\right)\right]
\cos \Theta (z, t)
\eea
where $D_0=\frac{\hbar A_0 g}{|{\bf e}_p\cdot{\bf p}_{31}|t_{0}}\sqrt{\left|\frac{{\rm Re}(K_{2})}{{\rm Re}(W)}\right|}$, and
$\Theta (z, t)=(k_{p}+K_{0})z-\omega_{p}t+A_{0}^{2}g^{2}z/(8L_{\rm disp})$
. The part of the quantum fluctuations is given by
\begin{widetext}
\begin{align}
\hat{{\bf E}}_{\rm fluc}
&={\bf e}_p\frac{D_0}{\sqrt{A_0}} \sum_{j=1}^2 \left\{\left[u_j(t)\cos (\theta+\Theta)
+v_j(t)\cos (\theta-\Theta)\right]\hat{X}_{j,\theta}
-\left[u_j(t)\sin (\theta+\Theta)+v_j(t)\sin (\theta-\Theta)\right]\hat{X}_{j,\theta+\frac{\pi}{2}}\right\}.
\end{align}
\end{widetext}
We see that the quantum fluctuations are characterized not only by the quadrature operators $\hat{X}_{j,\theta}$ and  $\hat{X}_{j,\theta+\frac{\pi}{2}}$, but also by the eigenmode functions $u_j$ and $v_j$ contributed by the discrete spectrum, which is quite different from cases in absence of soliton where the quantum squeezing is contributed only by continuous-spectrum quantum fluctuations, which have been neglected here because they are much smaller than those by the discrete spectrum. In addition, once the probe light is squeezed, the atomic spin squeezing can be realized.

Based on the results obtained in the last subsection, it is easy to get the expressions of the variances of $\hat{X}_{j,\theta}$, given by
\begin{subequations}\label{XVariance}
\begin{align}
&\langle\hat{X}_{1,\theta}^2(s)\rangle=\frac{1}{2}+\frac{1}{8}g^4A_{0}^4s^2\sin^2\theta
+\frac{1}{2}g^2A_{0}^2s \sin \theta \cos \theta,\label{XVariance1}\\
&\langle\hat{X}_{2,\theta}^2(s)\rangle =\frac{1}{2}+\frac{1}{8}g^4A_{0}^4s^2\cos^2\theta
+\frac{1}{2}g^2A_{0}^2s \sin \theta \cos \theta.\label{XVariance2}
\end{align}
\end{subequations}
We see that the quadrature variances of the $\langle\hat{X}_{1,\theta}^2(s)\rangle$ and $\langle\hat{X}_{2,\theta}^2(s)\rangle$ depart from their vacuum value 1/2, which means that the probe field displays quantum squeezing due to the existence of the Kerr nonlinearity (characterized by the nonlinear parameter $g$) in the system.

\begin{figure}
\centering
\includegraphics[width=1\columnwidth]{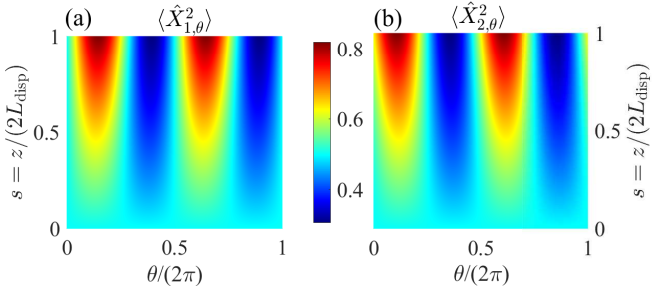}
\caption{Quantum squeezing of the slow-light soliton.
Quadrature variances $\langle\hat{X}_{j,\theta}^2\rangle$ ($j=1,2$) as functions of $s=z/(2L_{\rm disp})$ and $\theta/(2\pi)$ for $A_{0}=1$ and $g=1$. (a) and (b) are for $\langle\hat{X}_{1,\theta}^2\rangle$ and $\langle\hat{X}_{2,\theta}^2\rangle$, respectively.
Different colors shown in the color bar between the two panels denote different magnitudes of the quadrature variances. In the deep blue domains of $\theta$ and $s$, the quadrature variances are much smaller than their vacuum value, indicating that the slow-light soliton displays large quadrature squeezing.}
\label{Fig3}
\end{figure}

Shown in the panels (a) and (b) of Fig.~\ref{Fig3}
are respectively variances $\langle\hat{X}_{1,\theta}^2\rangle$ and  $\langle\hat{X}_{2,\theta}^2\rangle$ as functions of $s$ and $\theta$ by taking $A_{0}=1$ and $g=1$. One notes that when $s=0$ or $\theta=0$ both the variances take the vacuum values, i.e. $\langle\hat{X}_{1,0}^2(0)\rangle=\langle\hat{X}_{2,0}^2(0)\rangle=1/2$;
additionally, for any value of $s$ one has also $\langle\hat{X}_{1,0}^2(s)\rangle
=\langle\hat{X}_{2,\frac{\pi}{2}}^2(s)\rangle=1/2$.
However, when $\theta$ and $s$ locate in the deep blue domains of the both panels, the quadrature variances are much smaller than their vacuum value, which means that the soliton can indeed be quantum-mechanically squeezed in the atomic gas. The soliton can also be anti-squeezed, which occurs in the two deep red domains of the both panels. Furthermore, the result (\ref{XVariance}) shows that
the degree of squeezing (also antisqueezing) in the system gets larger during propagation (i.e. $s$ increases), which can also be seen clearly in Fig.~\ref{Fig3}. Since $\langle\hat{X}_{2,\theta}^2 (s)\rangle$ displays similar behaviors as $\langle\hat{X}_{1,\theta}^2 (s)\rangle$, in the following we discuss only  $\langle\hat{X}_{1,\theta}^2 (s)\rangle$ which characterize the quantum fluctuations of the amplitude and phase of the soliton.

\begin{figure}
\centering
\includegraphics[width=1\columnwidth]{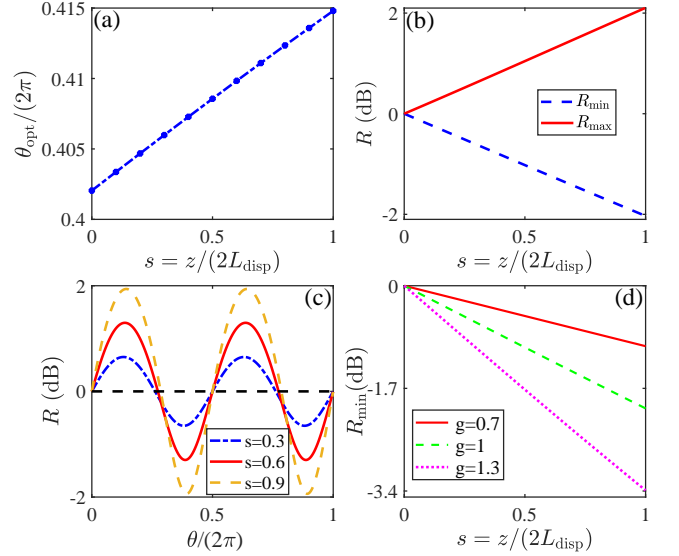}
\caption{(a)~Optimum angle $\theta_{\rm opt}$ for the quadrature variance $\langle\hat{X}_{1,\theta}^2\rangle$ as a function of propagation distance $s$.
(b)~Minimum squeezing ratio $R_{\rm min}$  (dashed blue line) and maximum squeezing ratio $R_{\rm max}$  (solid red line) versus $s$ (in unit dB).
(c)~Squeezing ratio $R$ versus angle $\theta$ with $s =0.3, 0.6 , 0.9$, plotted by
dotted blue, solid red, and dashed yellow lines, respectively.
Panels (a)-(c) are plotted with the parameters $A_{0}=1$ and $g=1$.
(d) Minimum squeezing ratio $R_{\rm min}$ as a function of $s$ for $A_{0}=1$ and different nonlinear parameter $g$. Solid red line: $g=0.7$; dashed green line: $g=1$; dotted pink line: $g=1.3$.}
\label{Fig4}
\end{figure}

By minimizing Eq.~(\ref{XVariance1}) with respect to $\theta$, we get the optimum angle as a function of the propagation distance $s$, i.e. $\theta_{\rm opt}=\theta_{\rm opt} (s)$, shown in Fig.~\ref{Fig4}(a)
for $A_{0}=1$ and $g=1$. With $\theta_{\rm opt} (s)$ we can get the minimum value of the quadrature as a function of $s$; meanwhile, the quadrature for the angle $\theta_{\mathrm{opt}}+\pi/2$  will be maximized. We can get an uncertainty ellipse in which $\langle\hat{X}^2_{1,\theta_{\rm opt}}(s)\rangle^{1/2}$ and $\langle\hat{X}^2_{1,\theta_{\rm opt}+\frac{\pi}{2} }(s)\rangle^{1/2}$ are along its minor and major axes, respectively.

Once $\theta_{\rm opt}(s)$ is known, experimentally one can choose the optimum detection angle to acquire the largest suppression of the quantum uncertainties in the amplitude and phase of the soliton.  One can define the squeezing ratio, i.e. the ratio of the quadrature variances between the value at position $s$ and that at the entrance $s=0$~\cite{HausJOSAB1990,YLai1993}
\begin{align}\label{R}
R=\frac{\langle\hat{X}_{1,\theta}^2(s)\rangle}{\langle\hat{X}_{1,\theta}^2(0)\rangle}
\end{align}
to characterize the degree of the squeezing quantitatively.

Fig.~\ref{Fig4}(b) shows the minimum squeezing ratio $R_{\rm min}$ and maximum squeezing ratio $R_{\rm max}$ as functions of $s$ (in unit dB), illustrated respectively by the dashed blue and the solid red lines.
We see that the quantum squeezing of the soliton found here is very efficient. This is due to the fact that the EIT-based atomic gas possesses giant Kerr nonlinearity and the soliton in such a system has an ultraslow propagating velocity, which makes the soliton have a significant squeezing only in a very short propagation distance (several centimeters). On the contrary, for acquiring the same degree of soliton squeezing in other systems (such as optical fibers), a much larger propagation distance is needed because of the weak Kerr nonlinearity and fast soliton propagation velocity in those systems.

Plotted in Fig.~\ref{Fig4}(c) is the squeezing ratio $R$ of the soliton as a function of angle $\theta$ for different propagation distance $s=0.3$ (dotted blue line), 0.6 (solid red line), and 0.9 (dashed yellow line), respectively.
One sees that the squeezing ratio is sensitive to the selections of $\theta$ and $s$. Fig.~\ref{Fig4}(d) shows the minimum squeezing ratio $R_{\rm min}$ as a function of $s$ for $A_{0}=1$ and different nonlinear parameter $g$. Solid red ,
dashed green , and dotted pink lines are for $g=0.7$, $1$, and $1.3$, respectively. We see that the minimum squeezing ratio is strongly dependent on the nonlinearity parameter $g$, and it decreases rapidly as the propagation distance $s$ is increased.

\subsubsection{Atomic spin squeezing}

The Kerr nonlinearity can not only result in the quantum squeezing of the probe laser field (as discussed above), but also cause atomic spin squeezing in the system. To show this, we consider the atomic spin operators~\cite{Kitagawa1993,Wineland1994}
\bes
\bea
&& \hat{s}_{x}=\frac{1}{2}(\hat{\sigma}_{12}+\hat{\sigma}_{21}),\\
&& \hat{s}_{y}=\frac{1}{2i}(\hat{\sigma}_{12}-\hat{\sigma}_{21}),\\
&& \hat{s}_{z}=\frac{1}{2}(\hat{\sigma}_{11}-\hat{\sigma}_{22}),
\eea
\ees
which satisfy the commutation relation
$\left[\hat{s}_{l},\hat{s}_{m}\right]=i\epsilon_{lmn}\hat{s}_{n}$
($\epsilon_{lmn}$ is the Levi-Civita antisymmetric unit tensor).
To calculate the spin squeezing, we introduce the quadrature spin operator
\begin{align}\label{spin2}
\hat{s}_{\theta}&=\frac{1}{2}\left[\hat{\sigma}_{12}e^{-i\theta}+\hat{\sigma}_{21}
e^{i\theta}\right],\notag\\
&=\cos\theta\,\hat{s}_{x}+\sin\theta\,\hat{s}_{y},
\end{align}
and define the spin squeezing degree
\begin{align}\label{spde}
\xi^{2}={\rm min}_{\theta}\left(\frac{\langle\hat{s}_{\theta}^{2}\rangle-\langle\hat{s}_{\theta}
\rangle^{2}}{\langle\hat{s}_{z}\rangle/2}\right).
\end{align}

From the result given by Eq.~(\ref{SE}) and the relation between $\hat{S}_{\a\b}$ and $\hat{\sigma}_{\a\b}$ [see the definition (\ref{Sab})], we obtain
\begin{align}
\hat{s}_{\theta}\approx-\frac{g_p \sqrt{n_0}}{2\Omega_c}\left(S_{\Phi}+\hat{\cal S}_{\Phi}\right),
\end{align}
where $S_\Phi=A_0\sqrt{g}{\rm sech}\left[\frac{A_{0}g}{2t_{0}}\left(t-\frac{z}{V_{g}}\right)\right]\cos\Phi$, $\hat{\cal S}_{\Phi}
=\sqrt{A_0 g}\sum_{j=1}^{2}\left(u_{j}\hat{X}_{j,\Phi}+v_{j}\hat{X}_{j,-\Phi}\right)$, with $\Phi=\Theta+\wp-\theta$ and $\wp=(k_p-k_c)z-(\omega_{p}-\omega_c)t$.
We see that the atomic spin includes also a soliton part (denoted by $S_\Phi$).

\begin{figure}
\centering
\includegraphics[width=1\columnwidth]{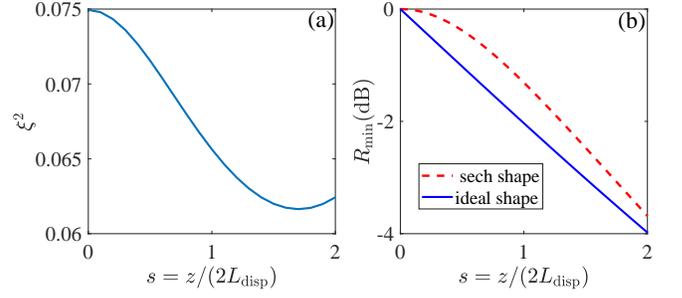}
\caption{(a)~Atomic spin squeezing degree $\xi^{2}$ as a function of propagation distance $s$. (b)~Minimum squeezing ratio $R_{\rm min}$ as a function of propagation distance $s$ for  $A_{0}=1$ and $g=1$.
The solid blue and dashed red lines are results for the LO pulse of the form $\varsigma_{j,\theta}(\sigma)$ and the sech-shaped LO pulse $\varsigma (\sigma)$, respectively.}
\label{Fig5}
\end{figure}

Shown in Fig.~\ref{Fig5}(a)
is the spin squeezing degree $\xi^{2}$ as a function of propagation distance $s$. One see that the system supports indeed atomic spin squeezing, which is also due to the existence of the Kerr nonlinearity in the system.

\subsubsection{Homodyne detection of the quantum squeezing}

Following the idea in Refs.~\cite{HausJOSAB1990,YLai1993}, we give a simple discussion on possible detection of the quadrature squeezing of the soliton. By using the Bogoliubov transformation (\ref{w}) and the orthonormal eigenfunctions given in the Appendix~(\ref{app3}), we can obtain
\begin{subequations}
\begin{align}\label{flu}
{\hat Q}_{j}(s)&=\frac{1}{\sqrt{2}}\int_{-\infty}^{+\infty}d\sigma \phi_{j}(\sigma)[{\hat w}(s,\sigma)+{\hat w}^{\dag}(s,\sigma)], \\
{\hat P}_{j}(s)&=\frac{1}{\sqrt{2}i}\int_{-\infty}^{+\infty}d\sigma \psi_{j}(\sigma)[{\hat w}(s,\sigma)-{\hat w}^{\dag}(s,\sigma)],
\end{align}
\end{subequations}
where $\psi_j(\sigma)=u_j(\sigma)+v_j(\sigma)$, $\phi_j(\sigma)=u_j(\sigma)-v_j(\sigma)$. Then, by using the definition (\ref{QuadOperators}) we have
\begin{eqnarray}\label{QD1}
&& {\hat X}_{j,\theta}(s)=\int_{-\infty}^{+\infty}d\sigma \left[\varsigma_{j,\theta}(\sigma)\hat{w}^\dag(s,\sigma)
+\varsigma_{j,\theta}^\ast(\sigma)\hat{w}(s,\sigma)\right],\nonumber\\
\end{eqnarray}
with
\be
\varsigma_{j,\theta}(\sigma)=[\cos \theta \phi_j (\sigma)+i\sin \theta \psi_j (\sigma)]/\sqrt{2}.
\ee
This result hints that one can employ a balanced homodyne detection technique to measure the quadrature variances by taking
$\varsigma_{j,\theta}(\sigma)$ as a coherent pulse injected into the atomic gas from a local oscillator (LO).
The measurement can be carried out as follows. First, the input probe pulse [i.e. the quantity $\hat{w}(s,\sigma)$ in (\ref{QD1})] is mixed with the LO pulse $\varsigma_{j,\theta}(\sigma)$ through a 50:50 beam splitter and the mixed signals from the two output paths are detected by two photodetectors respectively. Then, the difference of the output photo-currents from the two photodetectors is summed and integrated to complete the measurement. In fact, the expression (\ref{QD1}) can be understood as a projection of the input probe pulse $\hat{w}$ into the LO pulse through the measurement based on the homodyne detection.

Since reshaping a pulse to be a combination of some special functions [like
$\varsigma_{j,\theta}(\sigma)$ in (\ref{QD1})] is difficult, in practice it is more convenient to use a hyperbolic secant pulse, e.g. $\varsigma(\sigma)={\rm sech}(\sigma)\exp(i\theta )/\sqrt{2}$, as LO pulse. In this way,
the quadrature operator to be detected turns to be $\hat{X}_{1,\theta}'(s)= \cos\theta\,\hat{Q}_j+2\sin\theta\hat{P}_j$. Fig.~\ref{Fig5}(b)
shows the theoretical results on the minimum squeezing ratio $R_{\rm min}$ as a function of propagation distance $s$. In the figure, the solid blue and dashed red lines are ones for the LO pulse of the form  $\varsigma_{j,\theta}(\sigma)$ and the sech-shaped LO pulse $\varsigma (\sigma)$, respectively. We see that the minimum squeezing ratio obtained by using the hyperbolic-secant LO pulse is getting close to that obtained by the ideal LO pulse as propagation distance increases.

\section{Summary}\label{Sec5}

In this work, we have developed a quantum theory of slow-light solitons produced in an EIT-based atomic gas. Starting from the HLM equations which govern the quantum dynamics of the atoms and the quantized probe field, we have derived a quantum NLS equation controlling the evolution of the probe-field envelope. We have constructed an effective Hamiltonian and quantum Heisenberg equation of motion where the atomic variables have been eliminated. By exploiting a direct perturbation approach, we have diagonalized the effective Hamiltonian and carried out a detailed calculation on the quantum fluctuations of a slow-light soliton. These quantum fluctuations are expanded as a linear superposition of the complete and orthonormalized set of eigenfunctions obtained by solving the BdG equations. We have shown that, different from optical-fiber solitons, due to the giant Kerr nonlinearity contributed from the EIT effect, significant squeezing of the slow-light soliton can be obtained within a very short propagation distance both classically (i.e. squeezed soliton width) and quantum mechanically (i.e. squeezed quantum fluctuations). In addition, together with the the squeezing of the slow-light soliton, atomic spin squeezing can also be realized in the system. The results reported here is useful for understanding the quantum property of slow-light solitons and for realizing their quantum squeezing via EIT in cold atomic gases experimentally.

\section*{Acknowledgments}
This work was supported by the National Natural Science Foundation of China under Grant No.~11975098 and the Research Funds of Happiness Flower ECNU under Grant No.~2020ECNU-XFZH005.

\appendix

\section{Explicit expressions of the Heisenberg-Langevin equations}\label{app1}
Explicit expressions of the Heisenberg-Langevin equations (\ref{HLM}a) are given by
\begin{subequations}\label{HLMEs_explicit}
\begin{align}
& i\frac{\partial}{\partial t}{\hat S}_{22}-i\Gamma_{23}{\hat S}_{33}-\Omega_{c}{\hat S}_{23}+\Omega_{c}^{\ast}{\hat S}_{32}-i{\hat F}_{22}=0,\\
& i\left(\frac{\partial}{\partial t}+\Gamma_{3}\right){\hat S}_{33}+g_{p}{\hat S}_{13}\hat{E}_{p}-g_{p}^{\ast}\hat{E}_{p}^{\dag}{\hat S}_{31}+\Omega_{c}{\hat S}_{23}\nonumber\\
& -\Omega_{c}^{\ast}{\hat S}_{32}-i{\hat F}_{33}=0,\\
& \left(i\frac{\partial}{\partial t}+d_{21}\right){\hat S}_{21}+\Omega_{c}^{\ast}{\hat S}_{31}-g_{p}{\hat S}_{23}\hat{E}_{p}-i{\hat F}_{21}=0,\\
& \left(i\frac{\partial}{\partial t}+d_{31}\right){\hat S}_{31}+\Omega_{c}{\hat S}_{21}+g_{p}({\hat I}-{\hat S}_{22}-2{\hat S}_{33})\hat{E}_{p}\nonumber\\
&-i{\hat F}_{31}=0,\\
& \left(i\frac{\partial}{\partial t}+d_{32}\right){\hat S}_{32}+\Omega_{c}\left({\hat S}_{22}-{\hat S}_{33}\right)+g_{p}{\hat S}_{12}\hat{E}_{p}\nonumber\\
&-i{\hat F}_{32}=0,
\end{align}
\end{subequations}
and ${\hat S}_{11}={\hat I}-{\hat S}_{22}-{\hat S}_{33}$, with ${\hat I}$ the identity operator. In these equations, $d_{\alpha\beta}=\Delta_{\alpha}-\Delta_{\beta}+i\gamma_{\alpha\beta}$
($\alpha\neq \beta)$, $\gamma_{\alpha\beta}\equiv(\Gamma_\alpha+\Gamma_\beta)/2+\gamma_{\alpha\beta}^{\rm dep}$, and $\Gamma_\beta\equiv\sum_{\alpha<\beta}\Gamma_{\alpha\beta}$.
Here $\Gamma_{\alpha\beta}$ is the decay rate of spontaneous emission from the state $\beta$ to the state $\alpha$, $\gamma_{\alpha\beta}^{\rm dep}$ is the dephasing rate between $|\alpha\rangle$ and $|\beta\rangle$. ${\hat F}_{\alpha\beta}$ are $\delta$-correlated Langevin noise operators associated with the dissipation in the system, with the two-time correlation function given by
\begin{align}\label{FF}
\langle\hat{F}_{\alpha\beta}(z,t)\,\hat{F}_{\alpha'\beta'}&(z',t')\rangle=\notag\\
&\frac{L}{N}\delta(z-z')\delta(t-t'){\cal D}_{\alpha\beta,\alpha'\beta'}(z,t),
\end{align}
where ${\cal D}_{\alpha\beta,\alpha'\beta'}$ is atomic diffusion coefficient\cite{Kolchin}, which can be obtained from the Eqs.~(\ref{HLMEs_explicit}) using the generalized fluctuation dissipation theorem. Some of them are listed in the following
\begin{subequations}
\begin{align}
{\cal D}_{21,12}&=\Gamma_{23}\langle\hat{S}_{33}\rangle,\\
{\cal D}_{31,13}&=0,\\
{\cal D}_{\alpha1,1\beta}&=0~(\alpha,\beta=2,3;\alpha\neq\beta).
\end{align}
\end{subequations}

\section{Derivation of the QNLS equation~(\ref{QNLS0}) }\label{app0}

In the study on solitons in various classical physical systems, well developed reductive perturbation methods have been developed for simplifying complex, nonlinearly coupled partial differential equations into some well-known  ``amplitude equations'' (e.g. classical NLS equation{\color{red}~\cite{Huang2005PRE}}
, Korteweg-de Vries equation, etc.) that can be solved much easily~\cite{Jeffrey1982,Newell1992}. However, due to the difficulties for solving quantum nonlinear problems, up to now there is no quantum reductive perturbation method developed by which one can derive a QNLS equation directly from coupled nonlinear quantum partial differential equations involving many degrees of freedom of both atoms and quantized light fields.
In the following, we give a heuristic derivation on the QNLS equation describing the nonlinear evolution of the probe-field envelope $\hat{E}_{p}$ in the present system. The derivation of the QNSL equation can be divided into two steps.

{\it Step 1:  Quantum linear Schr\"odinger equation with group-velocity dispersion.}  We assume that the probe field is very weak so that the Kerr nonlinearity in the system can be neglected.
Thus the HLM equations (\ref{HLM}) can be treated by using a linear approximation.
By taking ${\hat S}_{\alpha\beta}\rightarrow {\hat S}_{\alpha\beta}^{(0)}+{\hat S}_{\alpha\beta}$, where ${\hat S}_{\alpha\beta}^{(0)}$ is the steady-state solution of ${\hat S}_{\alpha\beta}$ for $\hat{E}_{p}=0$ (i.e. ${\hat S}_{11}^{(0)}={\hat I}$, and other ${\hat S}_{\alpha\beta}^{(0)}=0$), we obtain
the linearized equations of Eqs.~(\ref{HLM}), which can be solved by using a Fourier transform. After eliminating the atomic variables, we obtain
\begin{align}\label{LSE}
\left[i\frac{\partial}{\partial z}+K(\omega)\right]{\tilde{\hat E}}_{p}(z,\omega)=i{\tilde{\hat{{\cal F}}}}_p(z,\omega).
\end{align}
Here $\omega$ is the sideband frequency of the probe pulse and $K$ is the linear dispersion relation defined by
\be
K(\omega)=\frac{\omega}{c}+\frac{|g_{p}|^{2}N}{c}\frac{\omega+d_{21}}{D(\omega)};
\ee
${\tilde{\hat E}}_{p}(z,\omega)$ and $\tilde{\hat{{\cal F}}}_p(z,\omega)$
are respectively the Fourier transforms of $\hat{E}_{p}(z,t)$ and  $\hat{\cal F}_p(z,t)$, i.e.
\begin{eqnarray}
&& {\tilde{\hat E}}_{p}(z,\omega)=\frac{1}{\sqrt{2\pi}} \int_{-\infty}^{\infty} dt {\hat E}_{p}(z,t)\,e^{-i\omega t},\\
&& \tilde{\hat{{\cal F}}}_{p}(z,\omega)=\frac{1}{\sqrt{2\pi}} \int_{-\infty}^{\infty} dt
\hat{{\cal F}}_{p}(z,t)\,e^{-i\omega t};
\end{eqnarray}
The new noise operator $\hat{{\cal F}}_{p}(z,t)$ is defined by
\be
\hat{{\cal F}}_{p}(z,t)=\frac{ g_{p}^\ast N}{c}\frac{\left(\omega+d_{21}\right) \hat{F}_{31}(z,t)-\Omega_{c}\hat{F}_{21}(z,t)}{D(\omega)},
\ee
where $D(\omega)=|\Omega_{c}|^2-(\omega+d_{21})(\omega+d_{31})$.

Assuming that the bandwidth of the probe pulse is not too narrow, one can expand $K(\omega)$ in a Taylor series around $\omega=0$ up to the second-order in $\omega$, i.e. $K(\omega)\approx K_{0}+\omega/V_{g}+K_{2}\omega^2/2$. Here $K_{0}\equiv K|_{\omega=0}$, $V_{g}^{-1}\equiv K_{1}\equiv(\partial K/\partial\omega)|_{\omega=0}$ is the group-velocity dispersion of the probe field, and $K_{2}\equiv(\partial^2K/\partial\omega^2)|_{\omega=0}$ is the coefficient denoting the group-velocity dispersion. Substituting this expansion into the envelope equation (\ref{LSE}) and convert it back to time domain by using an inverse Fourier transformation, we arrive the quantum linear Schr\"odinger equation
\begin{equation}\label{Linear Eq}
i\left(\frac{\partial}{\partial z}+\frac{1}{V_{g}}\frac{\partial}{\partial t}\right)\hat{E}_{p}+K_{0}\hat{E}_{p}-\frac{K_{2}}{2}\frac{\partial^2}{\partial t^2}\hat{E}_{p}-i{\hat{\cal F}}_{p}=0,
\end{equation}
where ${\hat{\cal F}}_p(z,t)$ is the inverse Fourier transform of ${\tilde{\hat{\cal F}}}_p(z,\omega)$.

{\it Step 2: Quantum nonlinear equation with cubic Kerr nonlinearity.} We next derive the equation for a weakly-nonlinear probe field for which the group-velocity dispersion can be neglected but the Kerr-nonlinearity is considered. This is valid when the probe pulse has a long-time duration, so that the time derivatives in the HLM Eqs.~(\ref{HLM}) play negligible roles.
To get the equation for $\hat{E}_{p}$ we employ an iteration method by taking
$g_{p}\hat{E}_{p}$ and a small quantity. Based on the steady-state solution  ${\hat S}_{11}={\hat I}$ and ${\hat S}_{22}={\hat S}_{33}=0$, we obtain the solution at the first-order approximation, given by
\be\label{SE}
{\hat S}_{\alpha1}=a_{\alpha1}^{(1)}g_{p}\hat{E}_{p}\,(\alpha=2,3)
\ee
and other ${\hat S}_{\alpha\beta}=0$, where
\begin{equation}\label{it1}
a_{\alpha1}^{(1)}=\frac{-\Omega_{c}^{\ast}\delta_{\alpha2}+d_{21}
\delta_{\alpha3}}{|\Omega_{c}|^2-d_{21}d_{31}}.
\end{equation}

Proceeding to the next order of iteration by substituting Eq.~(\ref{it1}) into
Eq.~(\ref{HLM1}) [i.e. Eq.~(\ref{HLMEs_explicit})], one obtains ${\hat S}_{\alpha\beta}=a_{\alpha\beta}^{(2)}|g_{p}|^2\hat{E}_{p}^{\dag}\hat{E}_{p}\,
(\alpha,\beta=1,2,3)$. Here
\begin{subequations}\label{it2}
\begin{align}
& a_{11}^{(2)}=\frac{\Gamma_{23}+2D_{c}}{\Gamma_{13}D_{c}}2{\rm Im}\left[a_{31}^{(1)\ast}\right]-\frac{1}{D_{c}}2{\rm Im}\left[\frac{\Omega_{c}^{\ast}}{d_{32}}a_{21}^{(1)\ast}\right],\\
& a_{22}^{(2)}=\frac{1}{D_{c}}2{\rm Im}\left[\frac{\Omega_{c}^{\ast}}{d_{32}}a_{21}^{(1)\ast}\right]
-\frac{\Gamma_{23}+D_{c}}{\Gamma_{13}D_{c}}2{\rm Im}\left[a_{31}^{(1)\ast}\right],\\
& a_{33}^{(2)}=-\frac{1}{\Gamma_{13}}2{\rm Im}\left[a_{31}^{(1)\ast}\right],\\
& a_{32}^{(2)}=-\frac{1}{d_{32}}\left[a_{21}^{(1)\ast}+\Omega_{c}
\left(a_{22}^{(2)}-a_{33}^{(2)}\right)\right],
\end{align}
\end{subequations}
and other ${\hat S}_{\alpha\beta}=0$, with $D_{c}=2\gamma_{32}|\Omega_{c}|^2/|d_{32}|^2$.

Based on the above results, we can proceed to the third-order of iteration. We get  ${\hat S}_{31}=a_{31}^{(3)}|g_{p}|^2g_{p}\hat{E}_{p}^{\dag}\hat{E}_{p}\hat{E}_{p}$, with
\begin{equation}\label{it3}
a_{31}^{(3)}=\frac{\Omega_{c}a_{32}^{(2)\ast}-d_{21}\left[a_{22}^{(2)}
+2a_{33}^{(2)}\right]}{|\Omega_{c}|^2-d_{21}d_{31}}.
\end{equation}
The solutions of other ${\hat S}_{\alpha\beta}$ are also obtained but are omitted here.

Exact to the third-order approximation with respect to $g_{p}\hat{E}_{p}$, we obtain the perturbation expansion of ${\hat S}_{31}$, given by
\begin{equation}\label{S31}
{\hat S}_{31}=a_{31}^{(1)}g_{p}\hat{E}_{p}+a_{31}^{(3)}|g_{p}|^2g_{p}
\hat{E}_{p}^{\dag}\hat{E}_{p}\hat{E}_{p}.
\end{equation}
Here the first (second) term on the right hand side of the above expression describes the linear (nonlinear) response of the atoms to the probe field.

Substituting Eq.~(\ref{S31}) into Eq.~(\ref{HLM}b), we arrive at the nonlinear equation
\begin{equation}\label{Nonlinear Eq}
\left(i\frac{\partial}{\partial z}+K_{0}\right)\hat{E}_{p}+W|g_{p}|^2\hat{E}_{p}^{\dag}\hat{E}_{p}\hat{E}_{p}=0,
\end{equation}
where $W=\hbar^2\omega_{p}\chi_{p}^{(3)}/(2c|{\bf e}_{p}\cdot\mathbf{p}_{31}|^{2})$ is the nonlinear coefficient contributed by the third-order Kerr-type nonlinearity, with
\be
\chi_{p}^{(3)}={\cal N}_{a}|{\bf e}_p\cdot{\bf p}_{31}|^4/(\varepsilon_{0}\hbar^3)a_{31}^{(3)}
\ee
the third-order nonlinear optical susceptibility (${\cal N}_{a}$ is atomic density).

By combining Eqs.~(\ref{Linear Eq}) and~(\ref{Nonlinear Eq}), we obtain the QNLS equation for $\hat{E}_{p}$:
\begin{align}\label{QNLS}
&\left[i\left(\frac{\partial}{\partial z}+\frac{1}{V_{g}}\frac{\partial}{\partial t}\right)+K_{0}\right]\hat{E}_{p}-\frac{K_{2}}{2}\frac{\partial^2}{\partial t^2}\hat{E}_{p}\notag\\
&+W|g_p|^2\hat{E}_{p}^{\dag}\hat{E}_{p}\hat{E}_{p}-i{\hat{\cal F}}_{p}=0,
\end{align}
which is valid for probe fields in which the group-velocity dispersion and cubic Kerr nonlinearity play equal roles. By making the transformation $\hat{E}_{p}\rightarrow\hat{E}_{p}\exp[i{\rm Re}(K_0)z]$, the above equation becomes the QNLS Eq.~(\ref{QNLS0}) given in the main text.

\section{Orthonormal and complete eigenfunctions of the linear eigenvalue problem on quantum fluctuations}\label{app3}

To solve the BdG Eqs.~(\ref{bdg}), we make the following transformation
\bea
&& u_q(\sigma)=\frac{1}{2}\left[\psi_q(\sigma)+\phi_q(\sigma)\right],\\
&& v_q(\sigma)=\frac{1}{2}\left[\psi_q(\sigma)-\phi_q(\sigma)\right],
\eea
$q=j, k$. Then we have the following equations
\begin{subequations}\label{EVP}
\begin{align}
& {\hat L}_{1}\psi_{j}(\sigma)=\lambda_{j}\phi_{j}(\sigma),
  \hspace{0.5 cm}   {\hat L}_{2}\phi_{j}(\sigma)=\lambda_{j}\psi_{j}(\sigma),\\
& {\hat L}_{1}\psi(\sigma,k)=\lambda_{k}\phi(\sigma,k),
     \hspace{0.1 cm}  {\hat L}_{2}\phi(\sigma,k)=\lambda_{k}\psi(\sigma,k),
\end{align}
\end{subequations}
with ${\hat L}_{j}=d^2/d\sigma^2+(2\delta_{j1}+6\delta_{j2}){\rm sech}^2(\sigma)-1$ ($j=1,2)$. Such equations have been considered by Yan {\it et al.} for the development of a direct perturbation theory on classical solitons~\cite{Yan1998PRE}, and later used for the study of the quantum solitons in Bose-Einstein condensates~\cite{Huang2006}.

Equations (\ref{EVP}) can be written as
\bes
\bea \label{EVP1}
&& \hat{L}_2\hat{L}_1\psi_q(\sigma)=\lambda_q^2 \psi_q (\sigma), \label{EVP11}\\
&& \hat{L}_1\hat{L}_2\phi_q (\sigma)=\lambda_q^2 \phi_q (\sigma),\label{EVP12}
\eea
\ees
where $q=j,k$; $\psi_k(\sigma)\equiv\psi(\sigma,k)$; $\phi_k(\sigma)\equiv\phi(\sigma,k)$.
Although both $\hat{L}_1$ and $\hat{L}_2$ are self-adjoint (Hermitian) operators, but $\hat{L}_1\hat{L}_2$ and $\hat{L}_2\hat{L}_1$ are not. However, since $(\hat{L}_2\hat{L}_1)^{\dag}=\hat{L}_1\hat{L}_2$, Eq.~(\ref{EVP11}) and Eq.~(\ref{EVP12}) are adjoint each other.
Solutions of the eigenvalue problem (\ref{EVP1}) are given by
\begin{subequations}
\begin{align}
& \phi (\sigma,k)=\frac{e^{ik\sigma}}{\sqrt{2\pi}(k^2+1)}\left[1-k^2-2ik\tanh(\sigma)\right], \\
& \psi(\sigma,k)=\frac{e^{ik\sigma}}{\sqrt{2\pi}(k^2+1)}
\big[1-k^2-2ik\tanh(\sigma)\notag\\
&\hspace{1.5cm}-2{\rm sech}^2(\sigma)\big],
\end{align}
\end{subequations}
for the continuous spectrum with eigenvalues $\lambda_k=-(k^2+1)$  $(-\infty <k<\infty)$, and
%
\begin{subequations}
\begin{align}
&  \phi_{1} (\sigma)={\rm sech}(\sigma), \\
&  \phi_{2} (\sigma)=\sigma{\rm sech}(\sigma),\\
&  \psi_{1} (\sigma)=\left[1-\sigma\tanh(\sigma)\right]{\rm sech}(\sigma), \\
&  \psi_{2} (\sigma)=\tanh(\sigma){\rm sech}(\sigma),
\end{align}
\end{subequations}
for the discrete spectrum with eigenvalues $\lambda_j=0$  ($j=1,2$).

The eigenfunctions of the discrete and continuous spectra given above satisfy the following orthonormal and complete relations
\begin{subequations}\label{orthonormality}
\begin{align}
& \int_{-\infty}^{+\infty}d\sigma \psi_{j}^{\ast}(\sigma) \phi (\sigma,k')=\int_{-\infty}^{+\infty}d\sigma \phi_{j}^{\ast}(\sigma) \psi (\sigma,k')=0,\\
& \int_{-\infty}^{+\infty}d\sigma \psi^{\ast}(\sigma,k) \phi (\sigma,k')=\int_{-\infty}^{+\infty}d\sigma \phi^{\ast}(\sigma,k) \psi (\sigma,k')\notag\\
&\hspace{3.76 cm}
=\delta(k-k'),\\
& \int_{-\infty}^{+\infty}d\sigma \psi_{j}^{\ast}(\sigma) \phi_{j'}(\sigma)=\delta_{jj'},\\
& \int_{-\infty}^{+\infty}dk \psi^{\ast}(\sigma,k) \phi(\sigma',k)+\sum_{j=1}^{2} \psi_{j}^{\ast}(\sigma) \phi_{j}(\sigma')\notag\\
&=\delta(\sigma-\sigma'),
\end{align}
\end{subequations}
i.e. $\{\psi_1(\sigma), \psi_2(\sigma), \psi ((\sigma,k)\}$, $\{\phi_1(\sigma), \phi_2(\sigma), \phi ((\sigma,k)\}$ are orthonormal and complete eigenfunctions forming a bi-orthonormal basis~\cite{Leung}.



\begin{references}
\bibitem{note}
If taking $R_0=200~\mu{\rm m}$ (the transverse radius of the probe field), one has the single-photon half Rabi frequency $g_{p}=1.14\times10^{6} ~{\rm s}^{-1}$. By the requirement $L_{\rm disp}=L_{\rm nln}=0.97$\,cm (i.e. $g\approx 1$) for forming soliton, we obtain $n_{0}=1.78\times10^{3}\gg 1$ using the system parameters indicated above.
\end{references}
\end{document}